\documentclass{aa}  

\usepackage{graphicx}
\usepackage[varg]{txfonts}
\usepackage[colorlinks=true,allcolors=blue]{hyperref}
\begin{document}

   \title{Making the unmodulated Pyramid wavefront sensor smart}
   \subtitle{Closed-loop demonstration of neural network wavefront reconstruction with MagAO-X}

   \titlerunning{Smart unmodulated PWFS}
   \authorrunning{R. Landman, S. Y. Haffert et al.}

   \author{
   R. Landman\inst{1}
   \and
   S.~Y.~Haffert\inst{2}
   \and 
   J.~R.~Males\inst{2}
    \and
   L.~M.~Close\inst{2}
   \and
   W.~B.~Foster\inst{2}
   \and
   K.~Van Gorkom\inst{2}
   \and
   O. Guyon\inst{2,3,4,5}
    \and
   A. Hedglen\inst{6}
   \and
   M. Kautz\inst{3}
   \and
   J. K. Kueny\inst{3}
   \and
   J. D. Long\inst{7}
   \and
   J. Lumbres \inst{3}
   \and
   E. A. McEwen \inst{3}
   \and
   A. McLeod \inst{8}
   \and
   L. Schatz\inst{9}
   }

   \institute{Leiden Observatory, Leiden University, PO Box 9513, 2300 RA Leiden, The Netherlands \\
              \email{rlandman@strw.leidenuniv.nl}
    \and
    Steward Observatory, The Unversity of Arizona, 933 North Cherry Avenue, Tucson, Arizona
    \and
    Wyant College of Optical Sciences, The University of Arizona, 1630 E University Blvd, Tucson, Arizona
    \and
    Subaru Telescope, National Observatory of Japan, National Institutes of Natural Sciences, 650 N. A'ohoku Place, Hilo, Hawai'i
    \and
    Astrobiology Center, National Institutes of Natural Sciences, 2-21-1 Osawa, Mitaka, Tokyo, Japan
    \and
    Northrop Grumman Corporation, 600 South Hicks Road, Rolling Meadows, Illinois
    \and
    Center for Computational Astrophysics, Flatiron Institute, 162 5th Avenue, New York, New York
    \and
    Draper Laboratory, 555 Technology Square, Cambridge, Massachusetts
    \and
     Starfire Optical Range, Kirtland Air Force Base, Albuquerque, New Mexico
    }

   \date{Received ; accepted}

 
  \abstract
  %
   {Almost all current and future high-contrast imaging instruments will use a Pyramid wavefront sensor (PWFS) as a primary or secondary wavefront sensor. The main issue with the PWFS is its nonlinear response to large phase aberrations, especially under strong atmospheric turbulence. Most instruments try to increase its linearity range by using dynamic modulation, but this leads to decreased sensitivity, most prominently for low-order modes, and makes it blind to petal-piston modes. In the push toward high-contrast imaging of fainter stars and deeper contrasts, there is a strong interest in using the PWFS in its unmodulated form. Here, we present closed-loop lab results of a nonlinear reconstructor for the unmodulated PWFS of the Magellan Adaptive Optics eXtreme (MagAO-X) system based on convolutional neural networks (CNNs). We show that our nonlinear reconstructor has a dynamic range of >600 nm root-mean-square (RMS), significantly outperforming the linear reconstructor that only has a 50 nm RMS dynamic range. The reconstructor behaves well in closed loop and can obtain >80\% Strehl at 875 nm under a large variety of conditions and reaches higher Strehl ratios than the linear reconstructor under all simulated conditions. The CNN reconstructor also achieves the theoretical sensitivity limit of a PWFS, showing that it does not lose its sensitivity in exchange for dynamic range. The current CNN’s computational time is 690 $\mu$s, which enables loop speeds of >1 kHz. On-sky tests are foreseen soon and will be important for pushing future high-contrast imaging instruments toward their limits.}


   \maketitle
%

\section{Introduction}
The direct imaging of extrasolar planets and circumstellar disks requires high contrast at small angular separations \citep{Bowler2016_imaging_review}. A crucial component of ground-based high-contrast imaging (HCI)
instruments is an extreme adaptive optics (AO) system that corrects for optical distortions from the atmosphere and the instrument itself \citep{guyon2018_xao_review}. To push HCI toward fainter stars and deeper contrasts, AO systems need highly sensitive wavefront sensors (WFSs) that optimally use all the photons to sense the incoming wavefront. Most current and future HCI instruments will use or are already using a Pyramid wavefront sensor \citep[PWFS;][]{Ragazzoni1996_pwfs} as their main WFS \citep[e.g.,][]{pinna2016soul, kasper2021pcs,males2022magao, males2022conceptual, haffert2022visible, bond2022_harmoni_ao}. Furthermore, upgrades of other HCI instruments are also planning to use a PWFS as a primary or second-stage WFS \citep{fitzsimmons2020gpi, perera2022gpi, boccaletti2022upgrading}. The increased preference for the PWFS over the Shack-Hartmann WFS is mostly due to its improved sensitivity \citep{Ragazzoni1999_pwfs_sensitivity, Chambouleyron2023A&A_noise_propagation_ffwfs}. This enhanced sensitivity allows AO systems to run at faster loop speeds, improving the achievable contrast at small angular separations. Additionally, its sensitivity can further be optimized by changing the binning fraction of the detector.

The main issue with the unmodulated PWFS is its nonlinear response to incoming wavefront aberrations, especially in the presence of strong turbulence \citep{Esposito2001_pwfs_partial_correction}. Without modulation, the linearity range of the PWFS is much smaller than the phase aberrations that are typically observed from atmospheric turbulence. As a result, most instruments use dynamic modulation to increase its linearity. However, this modulation reduces its sensitivity, especially to low-order modes, decreasing its performance on faint stars and limiting its loop speed. Additionally, it strongly reduces its ability to sense petal-piston modes, which are crucial for upcoming giant segmented telescopes \citep{Bertrou-Cantou2022_petalpiston_pwfs,Hedglen2022_segment_phasing_pwfs, Engler2022_flipflop}. Even when modulated, the PWFS still exhibits a nonlinear response to large aberrations. Furthermore, if a deformable mirror (DM) upstream of the PWFS is used to dig a dark hole or correct non-common path aberrations, the PWFS may need to operate with a nonzero offset, limiting its linearity range even more.

Many different solutions have been suggested to mitigate the nonlinearity of the PWFS. Some works have proposed a first-order correction to this nonlinearity in the form of an optical gain compensation \citep{Deo2019_optical_gain_tracking, Chambouleyron2020_opticalgain, Chambouleyron2021_opticalgain_focalplane}. However, this requires knowledge of the turbulence statistics and cannot account for nonlinear modal cross-talk. On the other hand, nonlinear reconstructors provide a software-based solution that can do higher-order correction without a reduction in sensitivity. Nonlinear reconstructors can be split into model-based and data-driven algorithms. Model-based methods rely on an accurate model of the optical system of the PWFS and accurate WFS to DM calibration. Model-based methods include reconstructors based on gradient descent \citep{Hutterer2018_hutterer_landweber, Frazin2018_gradient_based_estimation, Hutterer2023_nonlinear_reconstruction_maths} and the Gerchberg-Saxton algorithm \citep{Chambouleyron2023_gs_reconstruction}. These methods often need multiple iterations to converge to an accurate wavefront estimation. Alternatively, data-driven nonlinear reconstructors have shown promising results. These methods use the ability of neural networks (NNs) to approximate arbitrary functions to learn the inverse relation between WFS measurements and the incoming wavefront \citep{2018SPIE10703E..1FS_swanson_cnn_pred,Landman2020_nonlinear_cnn, Archinuk2023_nonlinear_reconstruction_ml, Wong2023_nn_reconstruction}. While these methods show promising results, there is a lack of closed-loop lab and on-sky demonstrations of these techniques. Finally, nonlinear or adaptive controllers may help alleviate some of the nonlinearity problems of the PWFS using time-domain information \citep{Landman2021_RL, 2021JATIS_wong_nonlinear_predictivecontrol,deo2021_correlation_locking_filter, haffert2021_ddspc, Nousiainen2022_rl, Pou2022_rl_mlreconstruction}. 

\begin{figure*}[htbp]
    \centering
    \includegraphics[width=\linewidth]{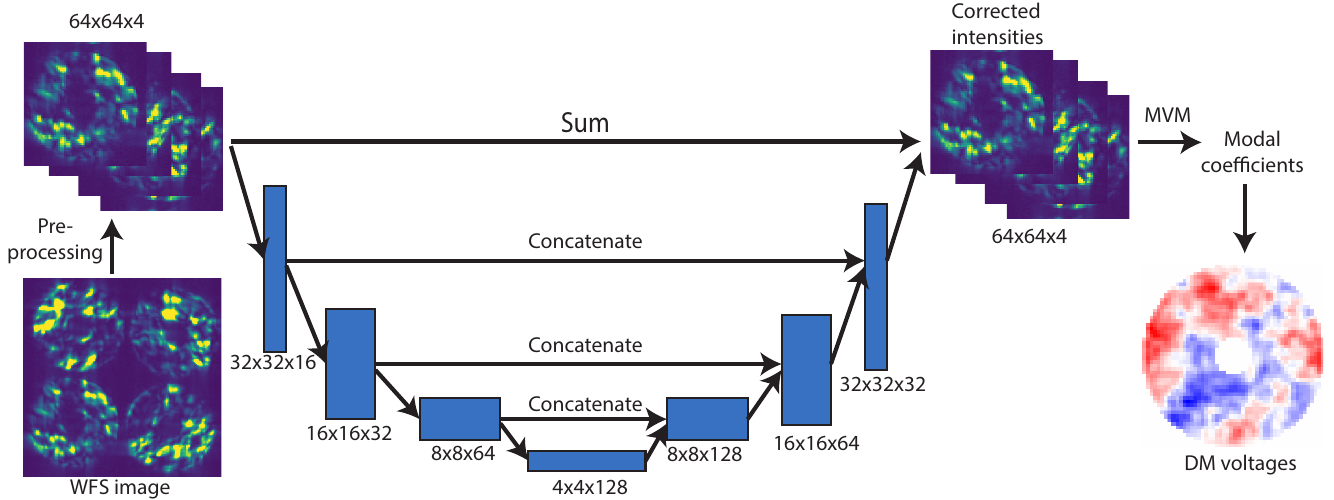
    }
    \caption{Visualization of the reconstruction pipeline and our NN architecture. The model consists of a U-net architecture with a nonlinear encoder and a decoder with skip connections. The nonlinear part uses leaky ReLU activation functions. Connections in the encoder part of the network use convolutional layers with a stride of 2, while the decoder part consists of transpose convolutional layers to ensure matching dimensions. }
    \label{fig:architecture}
\end{figure*}

In this work we present lab-based closed-loop tests of a convolutional neural network (CNN) reconstructor for the unmodulated PWFS of the Magellan Adaptive Optics eXtreme (MagAO-X) system. Section \ref{sec:methods} discusses the calibration and architecture of the CNN reconstructor. Section \ref{sec:reconstruction} presents the open-loop performance of the reconstructor compared to a linear reconstructor, while Section \ref{sec:closed_loop} compares the performance in a closed loop. Finally, Section \ref{sec:conclusions} summarizes the results and lists our conclusions.

\section{Methods}\label{sec:methods}
\subsection{MagAO-X}\label{sec:magaox}
MagAO-X is an extreme AO instrument for the 6.5 meter Magellan Clay Telescope at Las Campanas Observatory (LCO) in Chile. The instrument is shipped back and forth between Steward Observatory in Tucson, Arizona, and LCO. Shipping the instrument back and forth allows us 100\% access to the instrument when it is in Tucson. All the experiments shown in this work were performed at Steward Observatory.

MagAO-X is split into two optical benches that are connected by a periscope relay. The upper bench has the telescope simulator that is fed by a single-mode fiber-coupled super continuum laser (SuperK from NKT photonics). This source goes through the telescope simulator that generates an f/11 beam, which is equal to focal ratio of the Magellan Telescope. A pupil mask is used in an intermediate pupil to create the exact aperture of the Magellan Telescope. This beam is injected into the instrument and passes through several optics on the upper optical bench. The most important ones are the DMs. MagAO-X uses a woofer-tweeter architecture with a ALPAO-97 DM as woofer and a Boston Micromachines 2K tweeter \citep{males2018magaox, close2018magaox, males2022magao}. The beam is relayed by the periscope mirrors to the lower optical table, which contains the PWFS and the science instrumentation. A beamsplitter directly after the periscope system splits the light off into two paths; one for the PWFS for wavefront sensing and one for the science instrumentation. The AO beam goes through some additional flat mirrors and is collimated onto a high-speed piezo modulator. This beam is focused on the MagAO-X pyramid prism \citep{schatz2018design}. A custom triplet camera lens collimates the beam onto an electron-multiplying CDD (OCAM2K camera). The four PWFS pupils are sampled by 56 pixels across the pupil with a separation of 60 pixels. The science beam is focused by an off-axis parabola that creates a f/69 beam onto the science cameras. The science cameras sample the point spread function (PSF) with 3 pixels per $\lambda/D$ at H$\alpha$, which is 5.98 mas pixel$^{-1}$ on-sky (Long et. al. in prep).

\subsection{Training data}\label{sec:data}
One of the most important components toward obtaining a data-driven nonlinear reconstructor is generating an appropriate training set. This training set needs to be representative of the data that are seen on-sky. This is not trivial, as the distribution of aberrations in closed loop depends on among others on the turbulence statistics, loop gain and noise propagation for different spatial frequencies. To perform well over a large variety of conditions in a closed loop, we generate phase screens with random power-law power spectral densities and root-mean-square (RMS) phase. We uniformly sampled power-law indices between -1 and -3 and RMS values uniformly distributed in log-space between 0.2 nm and 600 nm. This ensures that the dataset contains the full range of aberrations that are seen on-sky in a closed-loop setting, from the large aberrations when closing the loop to the small aberrations after convergence. These phase screens were then projected on the controllable modes of the tweeter. For the reconstruction, we chose to reconstruct 1000 modes in a Fourier basis. We only reconstructed and controlled 1000 modes in to allow for the simulated turbulence, which is introduced using the same DM, to include a larger number of modes than we control. This mimics the situation on-sky, where we have to deal with the fitting error and subsequent optical gain effects. The modal coefficients for the training set were obtained by projecting the phase screens onto this modal basis. In total, we generated 100,000 phase screens, which were used for training, validation, and testing in a 60\%, 15\%, and 25\% split. All the data were collected with the ND2 neutral density filter inserted, which reduces the flux level by a factor of 100. This creates a photon flux that is roughly equivalent to observing a $\sim0$th magnitude star (I-band) at 2 kHz with MagAO-X.

\subsection{Neural network}\label{sec:cnn}
Artificial NNs are arbitrary function approximators. They consist of multiple sequential matrix vector multiplications (MVMs) with a nonlinear activation function in between, enabling them to learn nonlinear input-output relations. To make the model fitting tractable one can use domain knowledge to limit the amount of free parameters. The most prominent example of this are CNNs, which use the assumption that features are local and translationally invariant to drastically reduce the amount of free parameters. These CNNs have been used extensively for image processing tasks over the last decade and have revolutionized many fields of science. Since we are also dealing with images here, we chose to use CNNs as our model architecture. The architecture used here closely follows the one used in \citet{Landman2022_wfs_optimization} and is based on U-net \citep{2015arXiv150504597R_unet}, which uses skip connections to reduce the vanishing gradient problem. 
The input consists of square images of 64x64 pixels around the center of the pyramid pupils concatenated along its depth, resulting in an input of 64x64x4, as visualized in Fig. \ref{fig:architecture}. These images were normalized by their total intensity and scaled by a fixed constant such that the input data are between 0 and 1. This passes trough the nonlinear part of the network, which consists of an encoder and decoder. The leaky ReLU activation was used for each of the layers in the nonlinear part of the model. The output of this nonlinear part is a correction on the intensity image, which is added to the original image. This is subsequently mapped to the modal coefficients with a standard MVM{, or a "fully connected" layer.} These modal coefficients were scaled by a fixed constant, given by the standard deviation in the training set. {The MVM here is optimized as part of the NN and is not the same as the linear reconstructor derived in Section \ref{sec:linear}.}
The main advantage of this architecture is the presence of a direct linear connection between the input image and the output modal coefficients, allowing it to mimic a linear reconstructor with ease. This integrates the linear term that was found necessary for stable closed-loop operation in \citet{Landman2020_nonlinear_cnn} and \citet{Pou2022_rl_mlreconstruction}.
The loss function $J$ used here is based on the relative loss used in \citet{Landman2020_nonlinear_cnn} and is given by the ratio of the residual RMS error divided by the input RMS, with a constant $\epsilon$ to avoid divergence for very small input RMS. The loss is defined as follows:

\begin{equation}
    J = \left<\frac{\sqrt{\sum_i (y_{\textrm{true}, i} - y_{\textrm{pred}, i})^2}}{\sqrt{\sum_i y_{\textrm{true}, i}^2}+\epsilon}\right>,
\end{equation}
where $<>$ denotes the mean over a sampled batch, $y_{\textrm{true,i}}$ the applied modal coefficients for mode $i$ and $y_{\textrm{pred,i}}$ the predicted modal coefficients by the NN for that mode. This loss ensures that the model focuses in equal amounts on the small and large aberrations. If we simply used the residual RMS as the loss, this would disproportionately focus on the large aberrations. This would lead to suboptimal closed-loop performance, as the residuals in closed loop are often small \citep{Landman2020_nonlinear_cnn}. Throughout this work we use $\epsilon = $ 2 nm, as this was found to be the limiting precision for the linear model. 

We found that this model was initially overfitting to the training data and we subsequently added regularization. We used Dropout \citep{srivastava2014a_dropout} layers after the first two convolutional layers with 10\% dropout and one with 30\% dropout before the MVM. Additionally, we used L$_2$ regularization with a value of $10^{-5}$ in the final MVM layer, as that layer has most of the free parameters. The Adam optimizer \citep{2014arXiv1412.6980K_adam} was used to train the CNN with an initial learning rate of 0.003, which was decayed after every epoch by a factor of 0.96, and a batch size of 32. Training of the model took about 1 hour on a single GeForce RTX 2080 Ti GPU.

\subsection{Linear model}\label{sec:linear}
The linear model was constructed using the same data that were used to train the CNN. However, we only included data with an RMS phase smaller than 20 nm to ensure we are in a regime where the WFS is mostly linear. We calculated the reduced PWFS intensities by subtracting the zero-point reference and dividing by the total intensity. The reconstruction matrix, $R,$ was then obtained using a regularized linear least squares:
\begin{equation}\label{eq:least_squares}
     R = (X^T X + \rho \mathbb{I})^{-1}X^T Y, 
\end{equation}
where $X$ is the matrix with in its rows the measured reduced intensities, $Y$ the matrix with in its rows the applied modal coefficients, $\rho$ the regularization parameter and $\mathbb{I}$ the identity matrix. We only included the first 14,000 measurements in the regression to make the inversion feasible, which was calculated using a singular value decomposition. The regularization parameter was optimized on a test dataset of 3,000 samples. This was done by trying different values of $\rho$ and choosing the one that gave the lowest reconstruction RMS on the test dataset. We found an optimal value of $\rho = 3 \times 10^{-4}$ for the regularization parameter. 

\begin{figure*}
    \centering
    \includegraphics[width=0.45\linewidth]{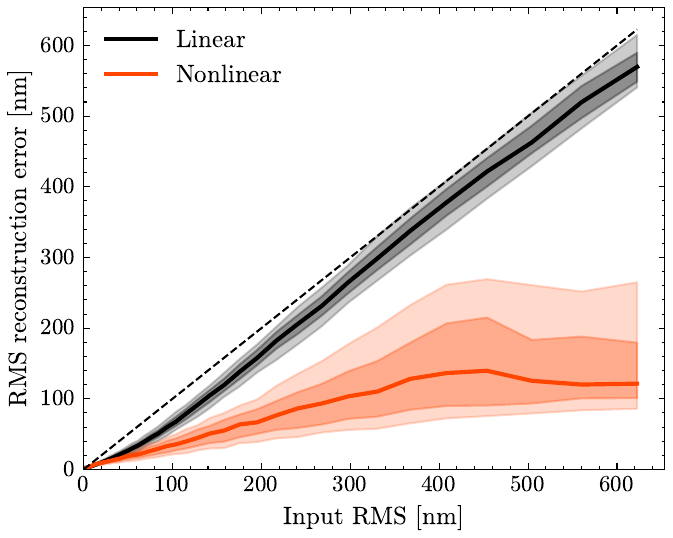}
    \includegraphics[width=0.45\linewidth]{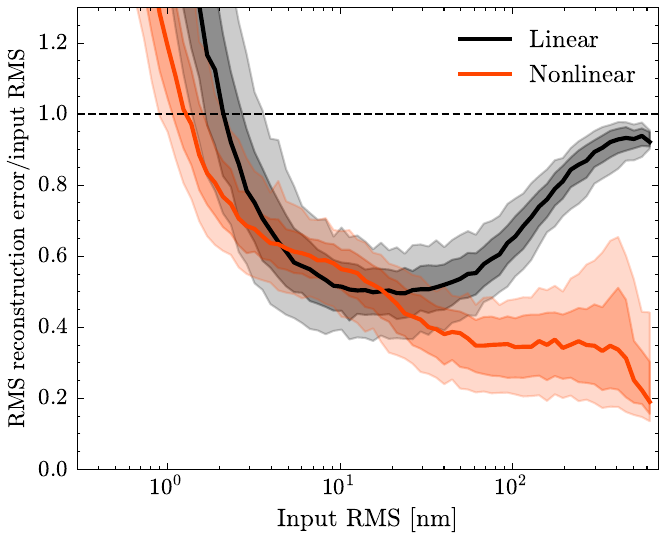}
    \caption{Comparison of the reconstruction accuracy of the nonlinear CNN with a linear model on the test dataset. The solid line indicates the mean value, and the colored region shows the 68\% and 95\% confidence intervals. The dotted line indicates where the RMS error of the reconstruction is equal to the input RMS, i.e., where there is no improvement. {Left:} Residual RMS error of the prediction as a function of the RMS of the input wavefront, shown on a linear scale. {Right}: Improvement factor, defined as the ratio between the residual RMS and the input RMS, as a function of the input RMS, which is shown on a logarithmic scale.}
    \label{fig:open_loop}
\end{figure*}

\section{Open-loop reconstruction}\label{sec:reconstruction}
\subsection{Reconstruction accuracy}
First, we compared the ability of the models to reconstruct the wavefront in an open-loop setting. To do this, we used the last 25\% of the generated data and evaluated the RMS error of the reconstruction, defined as
\begin{equation}
\textrm{RMS} = \sqrt{\sum_i (y_{\textrm{true}, i}-y_{\textrm{pred},i})^2}.
\end{equation}
The reconstruction RMS error as a function of the RMS of the input wavefront is shown in Fig. \ref{fig:open_loop}. We see that the NN has significantly reduced reconstruction RMS compared to the linear model. For the largest aberrations in our dataset (600 nm) there is a factor of 3 to 4 improvement in reconstruction accuracy with respect to the linear model, showing the increased dynamic range that can be obtained with a nonlinear reconstructor. The improvement factor, which is given by the residual RMS after reconstruction divided by the RMS of the input wavefront is shown on the right side of Fig. \ref{fig:open_loop}. When this curve intersects one the reconstruction does not lead to an improvement in wavefront quality. The intersection at the smallest input RMS therefore represents the sensitivity limit, while the intersection at the largest RMS determines the dynamic range. We observe that the NN is able to obtain a better sensitivity limit than the linear model. This is likely already due to nonlinearities, as the linear model was calibrated with wavefronts until 20 nm RMS, which is outside the fully linear regime of the unmodulated PWFS. The sensitivity limit for both reconstructors will be studied in more detail in the next section.

\begin{figure}[htbp]
    \centering
    \includegraphics[width=0.9\linewidth]{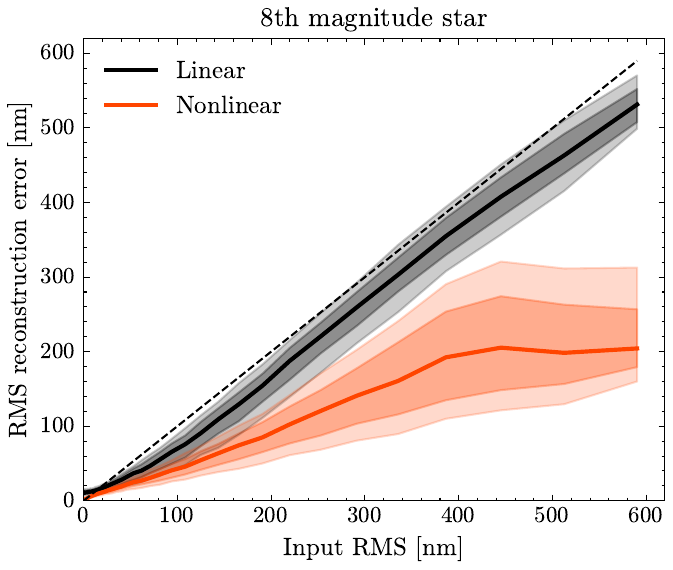}
    \caption{Comparison of the reconstruction accuracy for the nonlinear CNN with a linear model on the test dataset (same as Fig. \ref{fig:open_loop}), but now for an eighth magnitude star.}
    \label{fig:open_loop8th}
\end{figure}

\subsection{Reconstruction accuracy for fainter stars}
The main advantage of the unmodulated PWFS is its increased sensitivity, which allows for better AO performance on faint stars \citep{guyon2005limits,Agapito2023_nonmodulated_pwfs}. However, this performance benefit requires that the nonlinear reconstructor does not have stronger noise amplification than the linear model. Furthermore, in the case of noisy measurements, the NN may struggle to resolve nonlinear structure in the data, as this can be washed out by noise. To test the performance of the reconstructor on noisy data, we artificially increased the noise level of the test data and evaluated its open-loop reconstruction performance. This was done by adding Poisson noise for a given stellar magnitude. We ignored read noise as photon noise dominates for most WFSs using modern electron-multiplying CCDs, especially for the relatively bright natural guide stars in HCI. We converted the stellar magnitude to a total photon flux using the known zero point I-band magnitude for MagAO-X for the WFS arm \footnote{MagAO-x filter throughputs https://magao-x.org/docs/handbook/observers/filters.html}.

We then used transfer learning to convert the model used in the previous section to one that can deal with these noisy measurements. This was done by training the model on the same dataset again, but now randomly sampling stellar magnitudes between 0 and 10. We then evaluated the open-loop reconstruction performance at the noise level for an eight magnitude star.

{The optimal linear reconstructor also depends on the noise properties of the WFS measurements, and we therefore re-optimized the regularization strength for the linear model for this noise level.} We found that 20 nm was within the noise limit, so we had to increase the maximum RMS that was used for the linear regression to 100 nm. The results are shown in Fig. \ref{fig:open_loop8th}, showing a decrease in performance of the nonlinear reconstructor on these noisy data. This is likely because it starts to struggle to distinguish between nonlinear structure and noise. This is in good agreement with our previous results \citep{Landman2020_nonlinear_cnn}, in which we showed that the nonlinear correction gives the most improvement for bright stars. However, we still observe a major improvement over the linear model for large aberrations at this noise level.

\subsection{Sensitivity limit}
\begin{figure}[htbp]
    \centering
    \includegraphics[width=\linewidth]{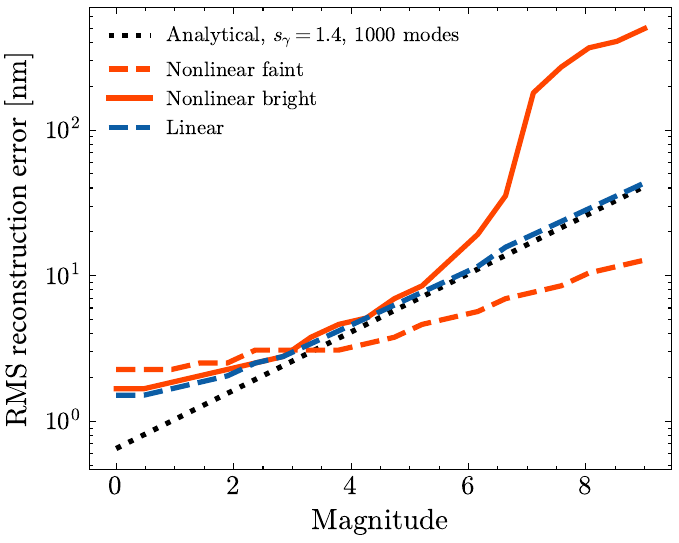}
    \caption{Smallest wavefront RMS that can be reconstructed as a function of the stellar I-band magnitude for the different reconstruction models. The analytical curve assumes a constant photon noise sensitivity of 1.4 for the unmodulated PWFS and the reconstruction of 1000 modes. The "Nonlinear bright" model refers to the model trained only on high signal-to-noise ratio data, while the "Nonlinear faint" model refers to the model that is also trained on noisy data.}
    \label{fig:sensitivity}
\end{figure}

To study the noise propagation properties of our reconstructor, we calculated the open-loop reconstruction performance for different stellar magnitudes. We then determined its sensitivity limit by finding the point at which the reconstruction RMS is $95\%$ of the input RMS, that is, the limit of the smallest aberrations that can still be reconstructed. The resulting RMS sensitivity limit as a function of stellar magnitude is shown in Fig. \ref{fig:sensitivity}. We show both the model trained on the noisy data ("Nonlinear faint") as well as the model only trained on the original high signal-to-noise ratio data ("Nonlinear bright"). The noise propagation of a Fourier-filtering WFS for a mode $\phi_i$ was derived in \citet{Chambouleyron2023A&A_noise_propagation_ffwfs}. If dominated by photon noise, this is given by
\begin{equation}
\sigma(\phi_i) =\frac{1}{s_\gamma^2(\phi_i) N_{ph}},
\end{equation}
where $\sigma(\phi_i)$ is the RMS for mode $\phi_i$, $s_\gamma$ the photon noise sensitivity of the WFS to that mode and $N_{ph}$ the total number of photons in the incoming beam. Assuming that the photon noise sensitivity for the unmodulated PWFS is roughly the same for each mode, the total RMS is given by\begin{equation}
    \sigma = \frac{1}{s_\gamma}\sqrt{\frac{N_{modes}}{N_{ph}}},
\end{equation}
where $N_{modes}$ denotes the number of reconstructed modes. The unmodulated PWFS has $s_\gamma \approx 1.4$ \citep{guyon2005limits, Chambouleyron2023A&A_noise_propagation_ffwfs}. The curve for this theoretical performance, assuming 1000 modes are reconstructed, is also shown in Fig. \ref{fig:sensitivity}. Additionally, we recalibrated the linear model {on the high signal-to-noise ratio data} by only including input wavefronts up until 6 nm. This avoids the linear model being limited by nonlinearities and allows us to really obtain the sensitivity limit. Fig. \ref{fig:sensitivity} shows that for the brightest stars we are limited by systematics caused by bench turbulence, which is why the curve flattens of to a limit of $\sim 1.5$ nm RMS. We see that both the linear model and the bright nonlinear model follow the same trend until a magnitude of $\sim$5, after which the CNN has much stronger noise amplification than the linear model. On the other hand, the nonlinear model trained on noisy data has better performance than the linear model and appears to even improve upon the analytical performance. We also see that it does not follow the classical $\propto 1/\sqrt{N_{ph}}$ curve. This is likely the result of the intrinsic regularization that is obtained by training the NN on noisy data. This effectively leads to the CNN being able to adapt the number of modes it reconstructs based on the WFS measurement. Furthermore, it can result in a biased estimator that, for example, always slightly underestimates noisy high-order modes, thereby decreasing the noise propagation for these modes. The more advanced (nonlinear) regularization methods may help NNs deal with these noisy measurements, as was also noted in \citet{2021JATIS_wong_nonlinear_predictivecontrol}. This added regularization comes at the cost of slightly decreased reconstruction accuracy for the brightest stars. 

\section{Closed-loop tests}\label{sec:closed_loop}
We evaluated the performance of the NN in closed loop for a variety of observing conditions. For all of these tests, we used a leaky integrator as our controller with a global gain ($g$) of 0.4 and leakage ($l$) of 1\%, which are typical value used for MagAO-X on-sky, and the CNN only trained on the high signal-to-noise ratio data. For each iteration, the DM voltages were updated as follows:
\begin{equation}
 \textrm{DM}_{t+1} = (1-l) \textrm{DM}_t + g y_ \textrm{pred}
.\end{equation}
Additionally, we used a separate tip-tilt loop as we observed small drifts in long-term tests. This was likely the result of tip-tilt drifts during the acquisition of the training data and not explicitly using tip-tilt in our modal basis. We used a linear reconstructor for this tip-tilt loop with the total normalized intensity in each of the four pupils as inputs . This separate tip-tilt loop was run with a gain of 0.2 and leakage of 1\%.

\begin{figure*}[htbp]
    \centering
    \includegraphics[width=0.8\linewidth]{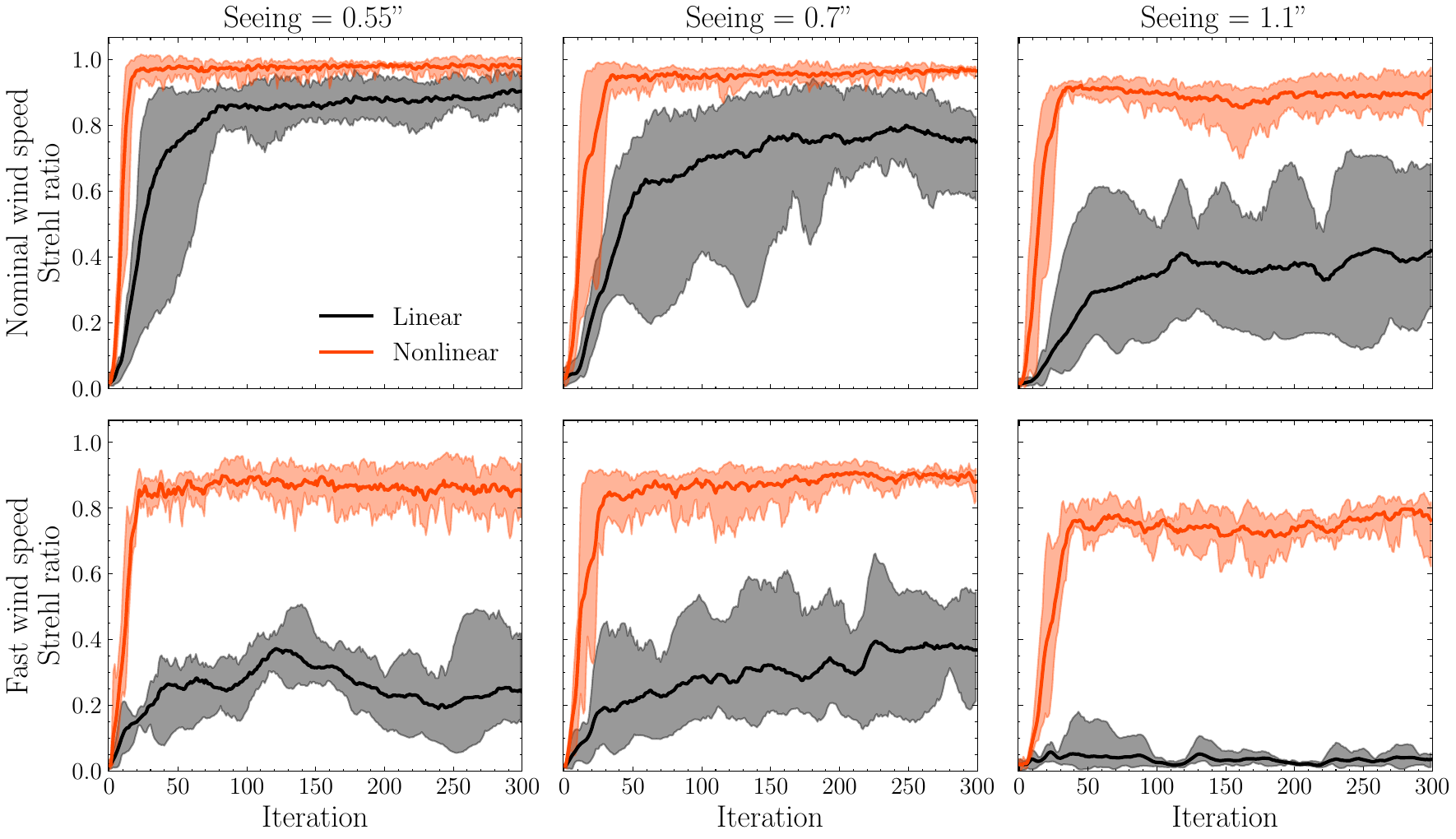}
    \caption{Closed-loop performance of the linear and nonlinear reconstructor for different atmospheric conditions. The solid line shows the mean Strehl at 875 nm over the five experiments, while the filled region indicates the minimum and maximum Strehl over the experiments. The top row shows the performance for a wind speed corresponding to median conditions at LCO with a $C_n^2$-weighted value of 18 ms$^{-1}$, while the bottom row uses double that value. The different columns correspond to different values for the seeing of the simulated turbulence. Videos of the closed-loop tests are available at \url{https://zenodo.org/records/10580651}.}
    \label{fig:strehl_curves}
\end{figure*}

We generated atmospheric turbulence with statistics similar as expected at LCO using hcipy \citep{por2018hcipy} and the data from \citet{prieto2010giant, thomas2010giant, males2018ground}, and projected this on the modes spanned by the tweeter. This means higher order modes that cannot be produced by the tweeter are not present in these tests. Still, this includes a bit over 1000 more modes than the 1000 that we are controlling, indicating that our model is not influenced by the presence of higher-order modes in closed loop. We ran tests for seeing values of 0.55, 0.7 and 1.1 arcseconds, which correspond to slight better than median, slightly worse than median and 75\% percentile conditions at LCO. As mentioned before, the tweeter cannot reproduce full atmospheric turbulence because of the limited number of modes. The fitting-error-limited Strehl using first-order estimates \citep{hardy1998adaptive} for these cases are 95\%, 92\%, and 85\%, respectively. We neglected the effect of the actuator influence function during the projection. The shape of the DM influence functions reduces the amplitude of higher-order modes \citep{ruane2020microelectromechanical}, and this creates slightly better seeing conditions than expected. We also tested the impact of the effective wind speed on the closed-loop performance by testing two different values, the median wind speed with an $C_n^2$ weighted value of 18 ms$^{-1}$ and double the median value at 36 ms$^{-1}$. While the tests were run using Python and not in real-time, we assumed an effective loop speed of 2 kHz in the simulation of the turbulence, which is the default loop speed for MagAO-X on bright stars \citep{males2022magao}. We repeated each of the experiments five times in order to test the stability and variance in performance.

The Strehl ratio was estimated from the focal plane images taken with the MagAO-X science camera. All Strehl measurements in this manuscript were done with the MagAO-X CH4 narrowband filter that has a center wavelength of 875 nm and a bandwidth of 26 nm \footnote{https://magao-x.org/docs/handbook/observers/filters.html}.  We did this by calculating the encircled energy within a radius of 1 $\lambda/D$ around the PSF peak. We normalized this by a reference Strehl measurement, which we obtained by closing the loop without applying any turbulence. The Strehl ratios shown throughout this work are relative to this reference measurement. We also found that the internal source had small drifts over time. To correct for this, we fitted a second order polynomial to the total intensity in the images over time and subsequently normalized the images. The resulting Strehl ratio curves are shown in Fig. \ref{fig:strehl_curves}. We observe an increase in Strehl ratio by using the nonlinear reconstructor over the linear reconstructor for each of the tested observing conditions. While the linear model is able to obtain $>80\%$ only in good seeing conditions, the nonlinear model is able to do this even with 1.1" seeing. The Strehl that is achieved by the nonlinear model is in good agreement with the fitting-error-limited estimates. The difference {between the reconstructors} becomes even larger in the case of strong winds, as shown in the bottom row of Fig. \ref{fig:strehl_curves}. The nonlinearity of the PWFS leads to an underestimation of the wavefront with a linear model. If the change of the wavefront in a single iteration is larger than this correction, it is stuck in the highly nonlinear regime and is unable to converge to a high Strehl. On the other hand, the improved estimation of large aberrations with the nonlinear reconstructor leads to higher convergence rates and a better estimation of the residuals in closed loop.

\subsection{Long-term stability}
The long-term stability of nonlinear models in closed loop cannot be guaranteed. For example, modes to which the WFS is not sensitive can start to creep onto the DM, requiring a lower gain or higher leakage to ensure long-term stability. Furthermore, if the model is overly sensitive to the WFS camera or pupil alignment, it may not be stable over multiple days and might require recalibration often. We found this to not be a big issue in our case. For example, the tests shown in the previous section were conducted two days after the collection of the data that were used to train the model. This might not seem very long but after power cycling MagAO-X, its pupil always has to be realigned on the tweeter and a separate linear stage must be used to align the PWFS pupil on the WFS camera. These steps are currently done by hand, which means there is some amount of randomness in the alignment every time the system is powered on. Being able to close the loop with the CNN several days after the data were taken shows that it is quite robust against misalignment. We therefore do not expect to need many recalibrations of the CNN model. {Even in the case of a required recalibration, we expect the recalibration to not take more than half an hour, as the previously calibrated model can be fine-tuned on newly collected data.}

\begin{figure}[htbp]
    \centering
    \includegraphics[width=\linewidth]{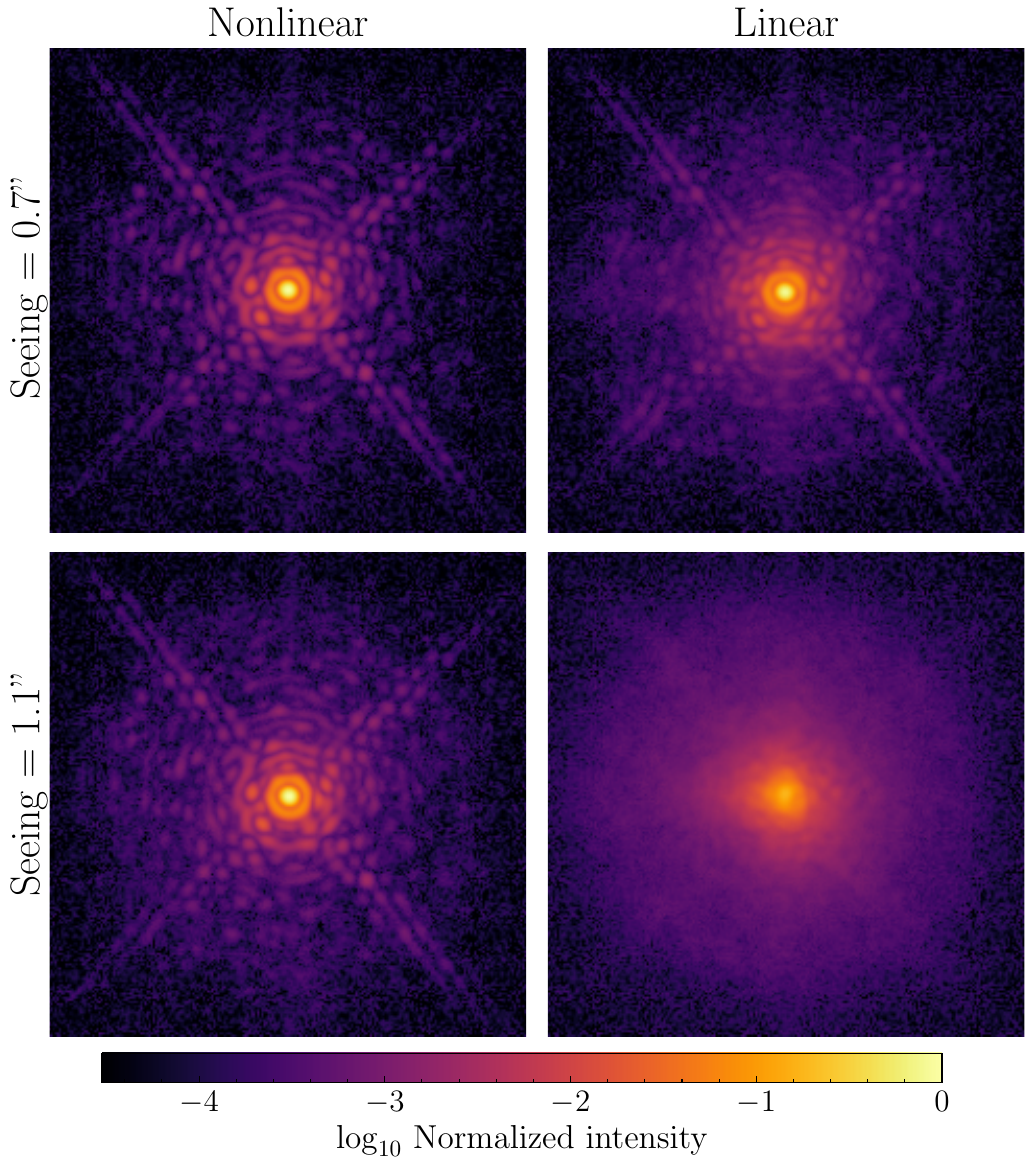}
    \caption{Integrated PSFs over a 2000-iteration closed-loop test for the nonlinear and linear reconstructor. The top row shows the performance for a seeing of 0.7", while the bottom row is for a seeing of 1.1". Videos of the closed-loop tests are available at \url{https://zenodo.org/records/10580651}.}
    \label{fig:longtermpsfs}
\end{figure}

To test its long-term stability, we ran a closed-loop test for 2000 iterations. We observed stable behavior of the PSF and no modal creep. The total integrated PSF for this test, while excluding the first 100 iterations, is shown in Fig. \ref{fig:longtermpsfs}. This shows the increased PSF stability with the nonlinear reconstructor as compared to the linear reconstructor, which will help boost exoplanet detection limits.

\begin{figure*}[htbp]
    \centering
    \includegraphics[width=0.8\linewidth]{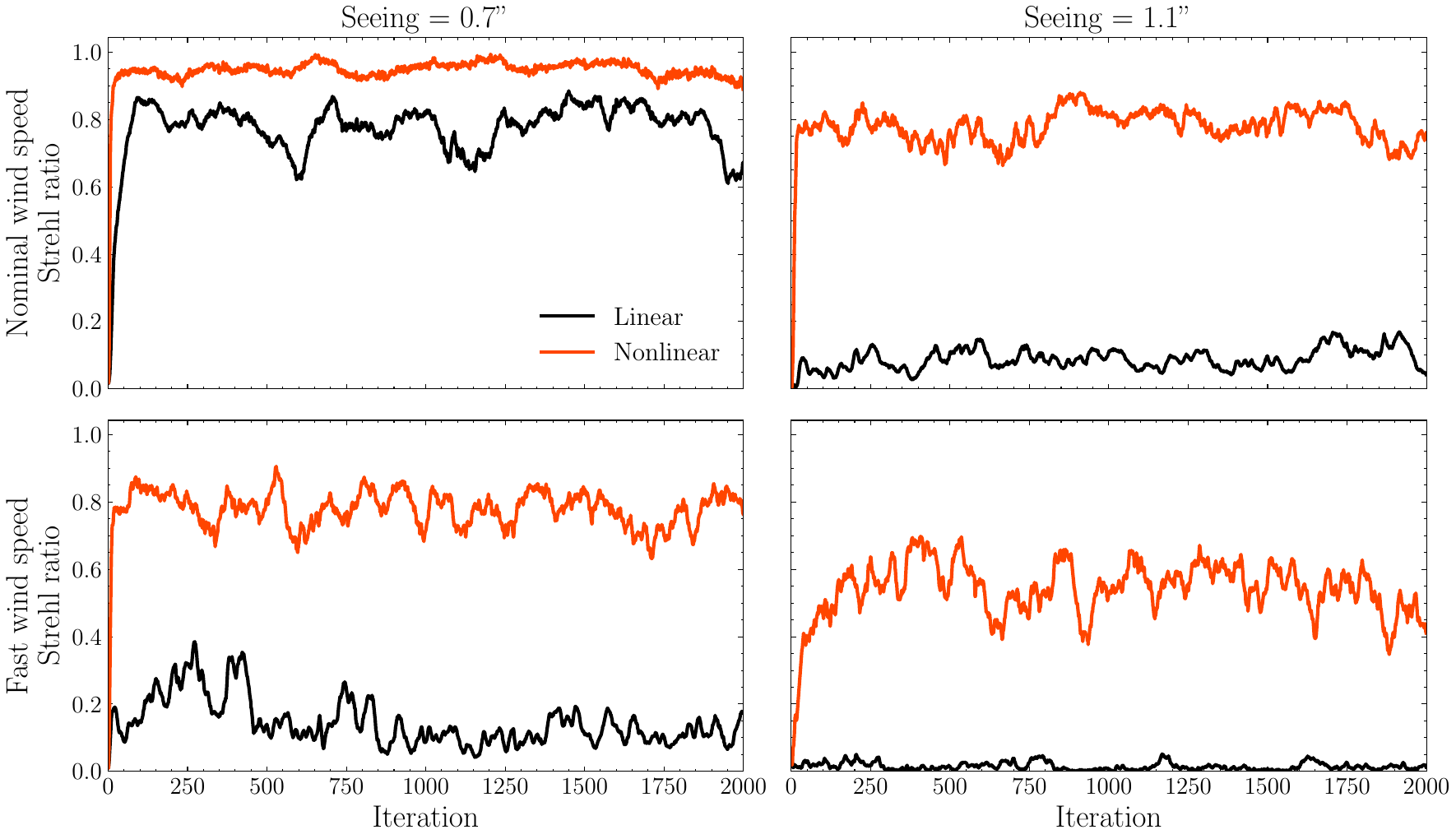}
    \caption{Closed-loop Strehl ratio for the linear and nonlinear reconstructors when the ND5 filter is used, decreasing the flux level by a factor of $10^3$ (or a $\Delta$ mag = 7.5) compared to the data that were used to train the models. The different rows correspond to different values for the wind speed, while the columns show the performance for different seeing values. Videos of the closed-loop tests are available at \url{https://zenodo.org/records/10580651}.}
    \label{fig:strehl_faint}
\end{figure*}

\subsection{Performance on fainter stars}
Next, we tried to close the loop on a flux level that the CNN was not trained on. While Fig. \ref{fig:sensitivity} shows that the models can obtain good reconstruction accuracy over a large range of flux levels, the closed-loop stability is not necessarily guaranteed. We evaluated this by testing the "bright" model, which was only trained on high signal-to-noise ratio data, in closed loop with the ND5 filter. This reduces the flux level by a factor of $10^{3}$ with respect to the training data and is equivalent to observing a $\sim$7.5th magnitude star with MagAO-X, similar to the flux level for Proxima Centauri.
We found that we did not have to tune the gain or leakage, as we observed stable behavior with the previously used values. The resulting Strehl ratios obtained during the closed-loop tests for different turbulence conditions is shown in Fig. \ref{fig:strehl_faint}. This shows a generally decreased performance of the AO system compared to the previous tests for both the linear and nonlinear model, which is due to the noisier WFS measurements. Still, the CNN has stable behavior in closed loop and can reach higher Strehl ratios than the linear reconstructor, showing that the nonlinear reconstructor can operate for a large range of stellar magnitudes. We note that Fig. \ref{fig:sensitivity} showed that the linear model has a better sensitivity limit for this stellar magnitude than the "bright" CNN, which we used for these test. However, the performance in closed-loop is still dominated by the nonlinearity error, as we are not operating around a diffraction-limited beam. This explains the improved performance of the CNN over the linear model, even though it has stronger noise propagation.

\subsection{Inference time}
Finally, we tested the inference speed of our model. We converted the trained model to a TensorRT optimized model with half precision, and tested the models on a single GeForce RTX 2080 Ti GPU. To remove the effect of overheads such as data transfer to the GPU on these measurements, we compared the inference speed of our models to an empty model, which returns a constant and does not do any processing. We find that reconstruction of a single WFS measurement takes on average $690 \pm 50$ $\mu s$, compared to $170 \pm 50$ $ \mu s$ for the linear model. This means that the inference is too slow to be run at 2 kHz, but loop speeds of >1 kHz are feasible. Pruning of the model could be used to further increase the inference speed \citep{Asif2019_pruning}. Alternatively, the size of the nonlinear model could be decreased. This comes at a slight decrease in reconstruction accuracy in the nonlinear regime, but in our experience this decrease is relatively small. Integration of TensorRT within the MagAO-X software environment, which is based on CACAO \citep{Guyon2018_cacao}, is currently ongoing. A true test of the inference speed and jitter of the model will be conducted after this integration and the model will be adjusted such that the standard 2 kHz loop speed can be achieved.

\section{Conclusions}\label{sec:conclusions}
We have presented the first closed-loop lab demonstration of a CNN-based reconstructor for the unmodulated PWFS with MagAO-X. Our nonlinear reconstructor has a significantly improved dynamic range of >600 nm RMS compared to the $\sim50$ nm RMS dynamic range for classical linear reconstruction. While its ability to accurately reconstruct the wavefront in the nonlinear regime decreases for fainter stars, we still observed a major improvement over a linear reconstructor. We have shown that the nonlinear reconstructor can reach the sensitivity limit of the PWFS and does not lead to stronger noise amplification when including noisy data in the training process. In this case, the nonlinear model does not obey the standard noise propagation scaling due to its intrinsic regularization properties. Closed-loop tests confirmed the increased dynamic range, with the nonlinear reconstructor reaching higher Strehl ratios than a classical linear reconstructor. The improved performance is the most pronounced when conditions are suboptimal, as the linear reconstructor is not able to converge in closed loop for the worst simulated conditions. We have extensively tested the stability of the nonlinear model and found that it is stable over multiple days, in long-term closed-loop tests and when tested on a different flux level than the one it was originally trained on. The presented work demonstrates that it is possible to use the unmodulated PWFS in most atmospheric conditions and that it might not be necessary to modulate the PWFS anymore in the future.

The obvious next step is to test the nonlinear reconstructor on-sky. Current work focuses on decreasing the computational complexity to reach 2 kHz speeds and implementing the TensorRT framework within the MagAO-X software environment to enable on-sky testing. This will make MagAO-X a pathfinder AO system for testing nonlinear control with (deep) NNs.

\begin{acknowledgements}
R.L. acknowledges funding from the European Research Council (ERC) under the European Union's Horizon 2020 research and innovation program under grant agreement No 694513. S.Y.H and the MagAO-X phase II project acknowledge generous support from the Heising-Simons Foundation. We are very grateful for support from the NSF MRI Award \#1625441 (MagAO-X).
\end{acknowledgements}

\bibliographystyle{aa}
\bibliography{references}

\end{document}